\documentclass[twocolumn,aps,prl,superscriptaddress]{revtex4-2}
\usepackage{amsmath}
\usepackage{graphicx}
\usepackage{MnSymbol}
\usepackage{hyperref}
\usepackage{fancyhdr}

\begin{document}

\title{Surface roughness induced stress concentration}

\author{B.N.J. Persson}
\affiliation{Peter Gr\"unberg Institute (PGI-1), Forschungszentrum J\"ulich, 52425, J\"ulich, Germany}
\affiliation{MultiscaleConsulting, Wolfshovener str. 2, 52428 J\"ulich, Germany}

\begin{abstract}
{\bf Abstract}: When a body is exposed to external forces large local stresses may occur at the surface because of surface
roughness. Surface stress concentration is important for many applications and in
particular for fatigue due to pulsating external forces. 
For randomly rough surfaces I calculate the probability distribution of surface stress in response to a
uniform external tensile stress with the displacement vector field parallel to the rough surface. 
I present numerical simulation results for the stress distribution $\sigma (x,y)$
and show that in a typical case, the maximum local tensile stress may be $\sim 10$ times bigger than the applied stress.
I discuss the role of the stress concentration on plastic deformation and surface crack 
generation and propagation. 
\end{abstract}

\maketitle

\thispagestyle{fancy}


{\bf 1 Introduction}

Almost all tribology applications involves surface roughness\cite{rev1,rev2,rev3,rev4,Carp}.
All surfaces of solids have roughness on many length scales\cite{MRS1,MRS2,MRS3,MRS4,MRS5}.
When an elastic body is deformed the stress at the surface will 
vary strongly with the position on the surface, and may take values much higher than the
applied stress, or the stress which would prevail if the surface would be perfectly smooth.
The points where the stress is high may act as crack nucleation centers.
In many practical applications bodies are exposed to deformations which fluctuate in time
and after many stress fluctuation cycles the body could breakup (fracture). This is denoted
as fatigue failure.

Silica glass is a good example for the influence of stress concentration at surface defects on its
tensile strength\cite{plast}. 
Imperfections of the glass, such as surface scratches, have a great effect on the strength of glass. 
Thus silica glass plates typically have a tensile strength of $\sim 7 \ {\rm MPa}$, 
but the theoretical upper bound on its strength is orders of magnitude 
higher: $\sim 17 \ {\rm GPa}$. This high value is due to 
the strong chemical Si-O bonds of silicon dioxide. 
The probability to find large defects decreases as the size of an object decreases. This is well known for
silica glass. Thus, glass fibers are typically 200-500 times stronger than for macroscopic 
glass plates.

Several different empirical 
equations have been presented for how to determine stress 
concentration at rough surfaces\cite{emp1,emp2}.
Two of them involve maximum height parameters such as $R_z$ (which I have denoted
$h_{1 z}$ in Ref. \cite{h1z}) which is determined mainly by the longest wavelength roughness
components (which have the largest amplitudes), and which will fluctuate strongly from
one measurement to another (see Ref. \cite{h1z,press}). However, most surfaces have self-affine fractal
surface roughness with a fractal dimension $D_{\rm f} > 2$ (or Hurst exponent $H<1$)\cite{PT,add2,add3}. 
In these cases the ratio between the height and the wavelength of a roughness component 
increases as the wavelength decrease, i.e., the roughness becomes sharper at short lengthscale. 
Thus, one cannot neglect the short wavelength roughness when calculating the stress concentration factor.

In this paper I will derive the probability distribution of surface stress 
for randomly rough surfaces, and show how it
can be used to estimate stress concentration factors. I will show that the maximum stress is proportional
to the root-mean-square (rms) surface slope.
I present numerical simulation results for height topographies and stress maps.
This is the third paper where I study statistical properties of randomly rough surfaces with applications.
The earlier papers focused on maximum height parameters with application to pressure fits\cite{press,h1z}.

\vskip 0.3cm
{\bf 2 Randomly rough surfaces}

All surfaces of solids have surface roughness, and many surfaces 
exhibit self-affine fractal behavior. This implies that if a surface area is magnified
new (shorter wavelength) roughness is observed which appears very similar to the roughness observed at smaller
magnification, assuming the vertical coordinate is scaled with an appropriate factor. 

The roughness profile $z=h({\bf x})$, where ${\bf x} = (x,y)$, 
of a surface can be written as a sum of plane waves ${\rm exp}(i{\bf q}\cdot {\bf x})$
with different wave vectors ${\bf q}$.
The wavenumber $q=|{\bf q}| = 2 \pi /\lambda$, where $\lambda$ is the wavelength of one roughness component.
The most important property of a rough surface is its power spectrum which can be written as
$$C({\bf q})  = {1\over (2\pi )^2} \int d^2 x \ \langle h({\bf x}) h({\bf 0})\rangle e^{i{\bf q}\cdot {\bf x}}\eqno(1)$$
where $\langle .. \rangle$ stands for ensemble averaging.
Defining 
$$h({\bf q})={1\over (2 \pi )^2} \int d^2 x \ h({\bf x}) e^{-i {\bf q}\cdot {\bf x}}\eqno(2)$$
$$h({\bf x})= \int d^2 q \ h({\bf q}) e^{i {\bf q}\cdot {\bf x}}\eqno(3)$$
one can show that (see Appendix A):
$$C({\bf q})  = {(2\pi )^2 \over A_0} \langle h({\bf q}) h(-{\bf q})\rangle \eqno(4)$$
where $A_0$ is the surface area.
Assuming that the surface has isotropic statistical properties,
$C({\bf q})$ depends only on the magnitude $q$ of the wave vector.
A self affine fractal surface has a power spectrum $C(q)\sim q^{-2(1+H)}$ (where $H$ is the Hurst exponent related to the 
fractal dimension $D_{\rm f} = 3-H$), which is a 
is a strait line with the slope $-2(1+H)$ when plotted on a log-log scale.
Most solids have surface roughness with the Hurst exponent $0.7 < H <1$ (see Ref. \cite{PT,add2,add3}).
 
For randomly rough surfaces, all the (ensemble averaged) information about the surface is contained in the 
power spectrum $C({\bf q})$. For this reason
the only information about the surface roughness which enter in contact mechanics
theories (with or without adhesion) is the function $C({\bf q})$.
Thus, the  (ensemble averaged) area of real contact, the interfacial stress distribution and the
distribution of interfacial separations, are all determined by 
$C({\bf q})$\cite{Persson2,Prodanov,Carbone1}.

Note that moments of the power spectrum determines standard
quantities which are output of most topography instruments and often quoted.
Thus, for example, the mean-square roughness amplitude 
$$h_{\rm rms}^2 = \langle h^2 \rangle =\int d^2q \ C({\bf q})\eqno(5)$$
and the mean-square slope 
$$\xi^2=\langle (\nabla h)^2\rangle = \int d^2q \ q^2 C({\bf q})\eqno(6)$$ 
are easily obtained as integrals involving $C({\bf q})$. We will denote the root-mean-square (rms) 
roughness amplitude with $h_{\rm rms}$
and the rms slope with $\xi$. If $C(q)$ denote the angular average (in ${\bf q}$-space) of $C({\bf q})$ then
from (6):
$$\xi^2 =  2 \pi  \int_{q_0}^{q_1} dq \ q^3 C(q). \eqno(7)$$
Assuming $C(q)=C_0 q^{-2-2H}$ this gives
$$\xi^2 =  2 \pi \int_{q_0}^{q_1} dq \ C_0 q^{1-2H}$$
For $H=1$ this gives
$$\xi^2 =  2 \pi  \int_{q_0}^{q_1} dq \ C_0 q^{-1}$$
If we write $q=q_0 e^\mu$ and $\mu_1= {\rm ln} (q_1/q_0)$ this gives
$$\xi^2 =  2 \pi \int_1^{\mu_1} d\mu \ C_0 $$
which shows that each decade in length scale contribute equally to the rms slope when
$H=1$. When $H<1$ the short wavelength roughness will be more important but in typical
application $H$ is close to $1$ and we will assume this in the numerical study presented in Sec. 6.

Surfaces of bodies of engineering interest, e.g., a ball in a ball bearing or a cylinder in a combustion engine, 
have always a roll-off region for small wavenumbers $q$,
because such bodies have some macroscopic shape, but are designed to be smooth at length scales smaller that
the shape of the body. In these cases the roll-off wavelength 
is determined by the machining process, e.g., by the size of 
the particles in sand paper or on a grinding wheel.
If the roll-off region matters in a particular application depends on the size of the relevant or studied surface area. 
Thus, if the lateral size $L$ is small the wavenumber $q=2\pi /L$ may be so large that it will fall in the region 
where the surface roughness power spectrum exhibit self-affine
fractal scaling, and the roll-off region will not matter.
We note that some natural surfaces, such as surfaces produced by brittle fracture, have fractal-like roughness 
on all length scales up to the linear size of the body.

        \begin{figure}[!ht]
        \includegraphics[width=0.20\textwidth]{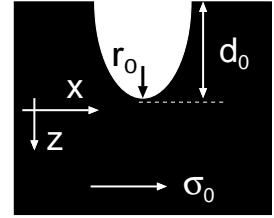}
\caption{\label{cavity.eps}
Half elliptic surface cavity (height $d_0$) in a rectangular solid block
exposed to the tensile stress which is uniform $\sigma_{xx} = \sigma_0$ far from the cavity.
The local stress close to the tip of the cavity is $\sigma_{xx} \approx \sigma_0 [1+2 \surd (d_0/r_0)]$
where $r_0$ is the radius of curvature at the cavity tip.
}
\end{figure}

        \begin{figure}[!ht]
        \includegraphics[width=0.33\textwidth]{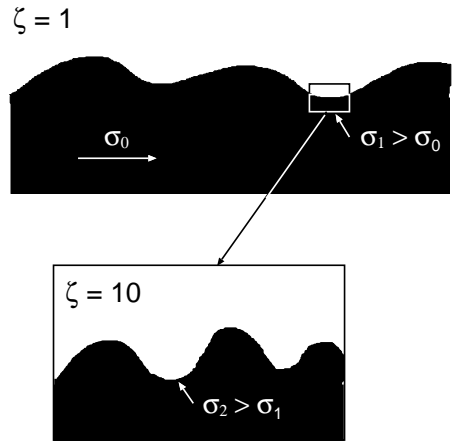}
\caption{\label{Many.eps}
Stress concentration: Surface roughness generate a local stress which is 
larger than the applied stress $\sigma_0$.
The local stress depends on the magnification $\zeta$ and increases as the magnification increases
because a ``cavity'' at the bottom of a bigger ``cavity'' experience already an enhanced
stress due to the larger cavity.
}
\end{figure}

\vskip 0.3cm
{\bf 3 Average stress concentration (approximate)}

Consider a half-elliptic cavity 
(height $d_0$ and radius of curvature at the bottom $r_0$) on the surface of 
a rectangular elastic block.
Assume that the block is elongated so the stress 
in the bulk far enough from the cavity is constant $\sigma_{xx}=\sigma_0$
while all other components of the stress tensor vanish. 
The local stress $\sigma_{xx}$ close to the tip of the cavity is denoted $S \sigma_0$, where 
the stress concentration factor (see Fig. \ref{cavity.eps}) (see Ref. \cite{notch,Pet,SK}) 
$S = 1+2\surd G$ where $G=d_0/r_0$. We are interested in the enhancement factor $S$ 
for randomly rough surfaces where there are
short wavelength roughness on top of longer wavelength roughness and so on,
where the qualitative picture presented in Appendix B prevail.

Here we will calculate the mean stress concentration by replacing $d_0/r_0$ with
$$G= - \langle h({\bf x}) \nabla^2 h ({\bf x})  \rangle \eqno(8)$$
where $\langle .. \rangle$ denotes ensemble averaging.. 
This approach includes the roughness on all length scales (see Fig. \ref{Many.eps}).
Since (8) is independent of the coordinate ${\bf x}$ we can average (8) over the $xy$-surface. Using (3) in (8) this gives 
$$G=  {1\over A_0} \int d^2x d^2q d^2q' \ q^2 \langle h({\bf q}) h({\bf q'})\rangle  
e^{i({\bf q}+{\bf q'})\cdot {\bf x}}$$
$$= {(2\pi )^2 \over A_0} \int d^2q \ q^2 \langle h({\bf q}) h(-{\bf q})\rangle \eqno(9)$$
where we have used that
$$\delta ({\bf q}+{\bf q'}) = {1\over (2\pi )^2} \int d^2x \ e^{i({\bf q}+{\bf q'})\cdot {\bf x}}$$
Using (4) we get from (9)
$$G=  \int d^2q \  q^2 C({\bf q}) = \xi^2$$
where $\xi$ is the surface mean slope. Thus 
$$S\approx 1+2 \xi . \eqno(10)$$

        \begin{figure}[!ht]
        \includegraphics[width=0.42\textwidth]{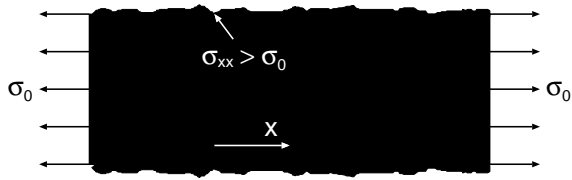}
\caption{\label{block.eps}
A rectangular block with surface roughness exposed to the elongation stress $\sigma_0$.
The surface roughness generate local stresses larger than the applied stress.
}
\end{figure}

\vskip 0.3cm
{\bf 4 Average stress concentration (exact)}

Assume that a rectangular block is elongated by the stress $\sigma_0$ (see Fig. \ref{block.eps}).
For a block with perfectly smooth surfaces the stress will be uniform in the block with
$\sigma_{xx} =\sigma_0$ and the other stress components equal to zero. When the block has
surface roughness the local stress at the surface could be much higher that the applied
stress $\sigma_0$ in particular at crack-like defects. 
Here we will calculate the rms stress concentration 
$$\sigma_{\rm rms} ^2 = \langle \left (\sigma_{xx}-\sigma_0 \right )^2 \rangle $$
We write
$$\sigma ({\bf x})= \sigma_{xx} ({\bf x})-\sigma_0$$
Using (3) and (see Appendix C)
$$\sigma ({\bf q}) = 2 \sigma_0 q f({\bf q}) h({\bf q})$$
where
$$f({\bf q}) = {q_x^2 \over q^2} \left (1+ \nu {q_y^2\over q^2} \right )$$
where $\nu$ is the Poisson ratio, gives
$$\langle \sigma^2 \rangle = {1\over A_0} \int d^2x \ \sigma^2 ({\bf x})$$
$$={2 \pi \over A_0} \int d^2q \ q^2 (2\sigma_0)^2 f^2 ({\bf q}) \langle h({\bf q}) h(-{\bf q}) )\rangle $$
$$=\int d^2q \ q^2 (2\sigma_0)^2 f^2 ({\bf q}) C({\bf q})\eqno(11)$$
If we assume roughness with isotropic statistic properties $C({\bf q})$ depends only on $q$
and in this case the angular integral in (11) can be performed analytically and we get (see Appendix C)
$$\langle \sigma^2 \rangle = (2\sigma_0)^2 \xi^2 g^2$$
where
$$g^2 = {1\over 8} \left (3+\nu +{3\over 16}\nu^2 \right )$$
Thus the rms stress concentration 
$$ \sigma_{\rm rms} = 2 \xi g \sigma_0 , \eqno(12)$$
In a typical case $\nu \approx 0.3$ giving $g \approx 0.64$ so (12) is consistent with (10).

\vskip 0.3cm
{\bf 5 Probability distribution of stress}

For an infinite system the probability distribution of stresses $\sigma = \sigma_{xx}-\sigma_0$ 
will be a Gaussian (see Appendix D):
$$P(\sigma) = {1\over (2\pi)^{1/2} \sigma_{\rm rms}} e^{-(\sigma/\sigma_{\rm rms})^2/2} , \eqno(13)$$
where $\sigma_{\rm rms} = 2 \xi g \sigma_0$. This equation
imply that there will be arbitrary high local stresses at some points. However, for any finite system the
probability to find very high stresses is small. We will now show how from (13) one can estimate the highest
stress at the surface. 

The stress probability distribution results from the fact that the stress is obtained by adding contributions 
to $P(\sigma)$ from each length scale with random phases. We have shown above that in a typical case where $H\approx 1$ each decade in 
length scale below the roll-off length scale gives approximately equal contributions to the 
rms slope and hence to $\sigma_{\rm rms}$.
Hence there will $N\approx (\lambda_{\rm r}/\lambda_1)^2$ important uncorrelated (because of the random phases) 
contributions to the probability distribution $P(\sigma)$ from the region $q_r < q <q_1$.
The roll-off region correspond to $(\lambda_0/\lambda_r)^2$ uncorrelated units so the total 
number of uncorrelated terms is $N\approx (\lambda_{\rm r}/\lambda_1)^2 (\lambda_0/\lambda_{\rm r})^2 = 
(q_1/q_0)^2$ which is the same as when no roll-off region exist. Note that this is very different
from the probability distribution for surface heights where the region $q>q_{\rm r}$ gives a fixed
number of uncorrelated terms independent of $q_1$ if $q_1/q_{\rm r} >> 1$.
The reason for this is that the height distribution $P(h)$ depends mainly on the
longest wavelength surface roughness components, which have the largest amplitudes.

An estimation of the maximum stress $\sigma_{\rm max}$ can be obtained from the condition
$$\int_{\sigma_{\rm max}}^\infty d\sigma \ P(\sigma) \approx N^{-1}\eqno(14)$$
Denoting $x=\sigma_{\rm max} /\sigma_{\rm rms}$ from (13) and (14) we get if $N>>1$:
$$x \approx \left [ 2 {\rm ln} \left ({N\over (2\pi)^{1/2} x} \right )\right ]^{1/2}\eqno(15)$$
In a typical case $\lambda_{\rm 0} =1 \ {\rm cm}$ and $\lambda_1 = 1 \ {\rm nm}$
giving $N=10^{14}$ and from (15)
$x\approx 7.7$ and the maximum stress is 
$\sigma_0+\sigma_{\rm max} \approx (1+15.4 \xi g) \sigma_0$. In a typical case
$\xi \approx 1$ and the maximum local stress will be $\sim 10$ times bigger than the applied stress.
We will consider the influence of plastic flow and crack formation on the 
roughness profile and the stress distribution in Sec. 7.

\begin{figure}
        \includegraphics[width=0.48\textwidth]{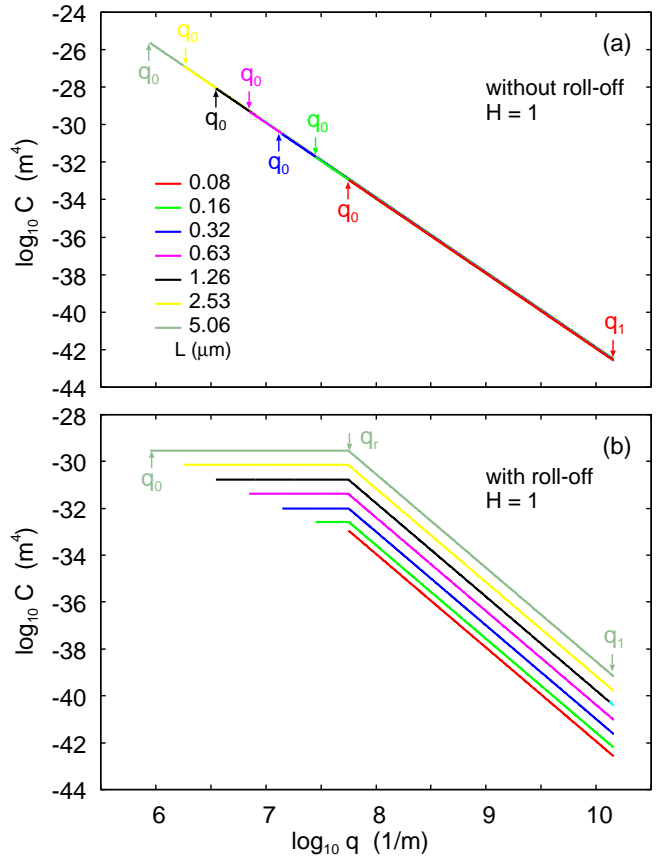}
        \caption{\label{1logq.2logC.with.and.without.roll.off.H=1.eps}
The surface roughness power spectra as a function of the wave number
(log-log-scale) used in the calculations of the surface height profile
for surfaces with the Hurst exponent $H=1$ without (a) and with (b) a roll-off region.
In (a) we indicate the large and small wavenumber cut-off $q_1$ and $q_0$, and the (b) also the
roll-off wavenumber $q_{\rm r}$. 
For each system size $L=2\pi/q_0$ the power spectra
have been chosen so the rms roughness amplitude $h_{\rm rms}$ are 
the same with and without the roll-off region.
}
\end{figure}

\begin{figure}
        \includegraphics[width=0.48\textwidth]{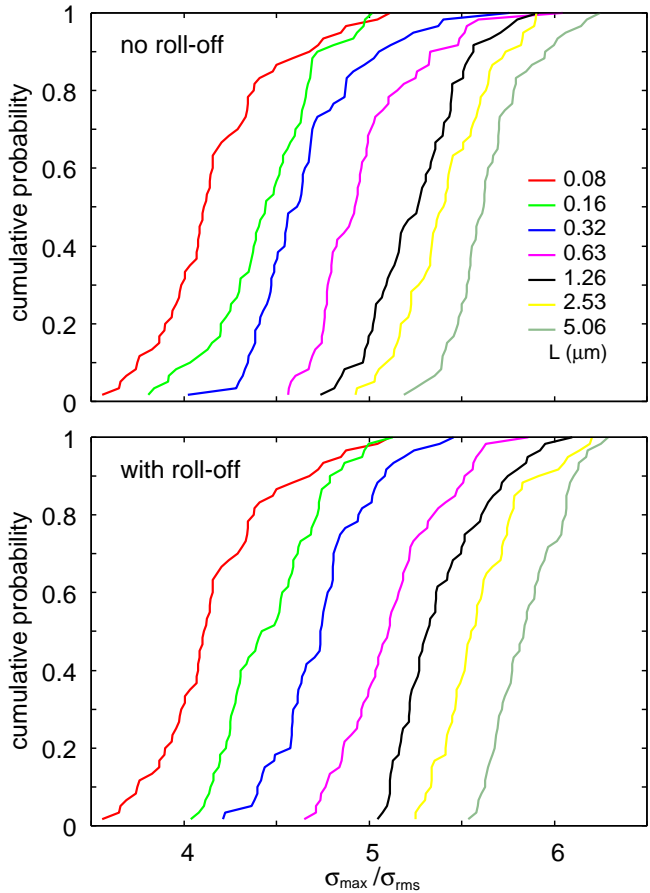}
        \caption{\label{1ratio.Pmax.over.prms.NoRollOff.eps}
The cumulative probability for the 
ratio $\sigma_{\rm max}/\sigma_{\rm rms}$ between the maximal surface stress and the rms surface stress
for the power spectra shown in Fig. \ref{1logq.2logC.with.and.without.roll.off.H=1.eps} 
without (a) and with (b) a roll-off region.
}
\end{figure}

\begin{figure}
        \includegraphics[width=0.48\textwidth]{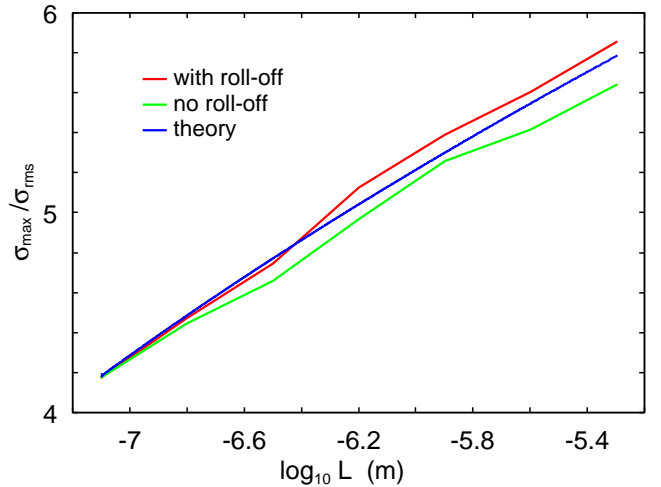}
        \caption{\label{1logLx.2ratio.sigMax.sigRMS.with.theory.eps}
The ratio $\sigma_{\rm max}/\sigma_{\rm rms}$ between the highest surface stress and the rms surface stress.
as a function of the logarithm of the size of the unit $L$. The red and green lines
are with and without a roll-off region in the power spectra, and the results are obtained after averaging over 60 realizations of the
surface roughness. The blue line is the theory prediction using (A3) with $N=(L/a)^2$ i.e. assuming that all the roughness components
contribute equally. 
}
\end{figure}

\begin{figure}
        \includegraphics[width=0.48\textwidth]{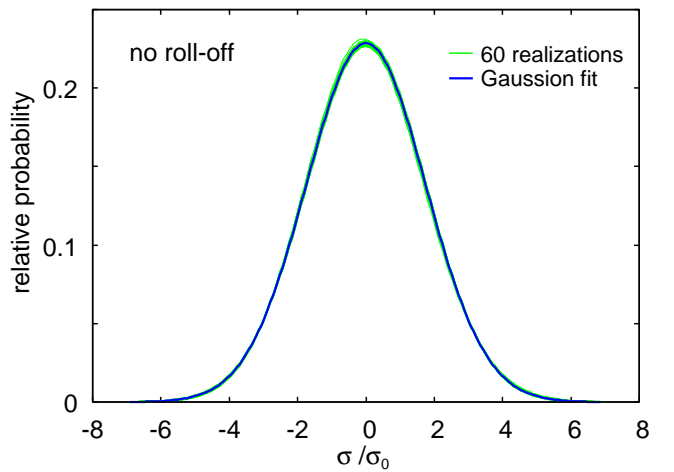}
        \caption{\label{1sigma.over.sigma0.2Probability.eps}
The stress probability distribution as a function of the stress as obtained from the computer simulations
using 60 realizations of the roughness (green lines) for the case without a roll-off in the
surface roughness power spectra. 
The blue line in is a Gaussian fit to the average of the green data. 
The calculations are for a surface with $L=2.53 \ {\rm \mu m}$ (with the rms slope $\xi = 0.8817$) using the
corresponding power spectra shown in Fig. \ref{1logq.2logC.with.and.without.roll.off.H=1.eps} (a) (Hurst exponent $H=1$).
}
\end{figure}

\vskip 0.3cm
{\bf 6 Numerical results}

We will now discuss the relation between the maximum surface stress $\sigma_{\rm max}$ 
and the rms stress $\sigma_{\rm rms} = 2 \xi g \sigma_0$.
No two surfaces have the same surface roughness, and $\sigma_{\rm max}$ 
will depend on the surface used. To take this into account we have generated surfaces
(with linear size $L$) with different random surface roughness but with the same surface roughness power spectrum.
That is, we use different realizations of the surface roughness
but with the same statistical properties. For each surface size we have generated
60 rough surfaces using different set of random numbers. The surface roughness
was generated as described in Ref. \cite{PT} (appendix A) by adding plane waves with random phases
$\phi_{\bf q}$ and with the amplitudes determined by the power spectrum:
$$h({\bf x}) = \sum_{\rm q} B_{\bf q} e^{i ({\bf q} \cdot {\bf x} + \phi_{\bf q})}\eqno(16)$$
where $B_{\bf q} = (2\pi /L) [C({\bf q})]^{1/2}$. We assume isotropic roughness so $B_{\bf q}$ and $C({\bf q})$ only depend on the
magnitude of the wavevector ${\bf q}$. The surface stress $\sigma_0+\sigma({\bf x})$ can be calculated from (C10) or 
can be generated directly using
$$\sigma({\bf x}) = \sigma_0 \sum_{\rm q} F_{\bf q} e^{i ({\bf q} \cdot {\bf x} + \phi_{\bf q})}\eqno(17)$$
where 
$$F_{\bf q} = 2q f({\bf q}) B_{\rm q} = (2\pi /L) \left [4 q^2 f^2({\bf q}) C({\bf q}) \right ]^{1/2} . \eqno(18)$$ 
In the present numerical study we will assume that the surface roughness has isotropic statistical properties 
so that $C({\bf q})$ only depends on $q=|{\bf q}|$. However, even in this case the stress $\sigma_{xx}$ 
has anisotropic statistical properties
because of the factor $f({\bf q})$ in (18). However, here we are only interested in comparing the prediction of (15) with the
numerical theory, and for this it is enough to replace $f$ with its angular average value $1/2+\nu /8$ which we can consider
as included in an effective $\sigma_0$. Thus we assume $f=1$ both in the numerical calculation 
and in (15) when comparing the theory with the numerical study.

We have used surfaces of square unit size, $L\times L$, with 7 different sizes, 
where $L$ increasing in steps of a factor of $2$ from
$L=79 \ {\rm nm}$ to $L=5.06 \ {\rm \mu m}$, 
corresponding to increasing $N$ from $N=256$ to $N=16384$. The lattice
constant $a \approx 0.309 \ {\rm nm}$. 

The longest wavelength roughness which can occur on a surface with size $L$
is $\lambda \approx L$ so when producing the roughness on a surface we only include the part of the power spectrum between
$q_0 < q < q_1$ where $q_0 = 2 \pi /L$ and where $q_1$ is a short distance cut-off corresponding to atomic dimension
(we use $q_1 = 1.4\times 10^{10} \ {\rm m^{-1}}$). 
This is illustrated in Fig. \ref{1logq.2logC.with.and.without.roll.off.H=1.eps} which shows the
different short wavenumber cut-off $q_0$ used. 

We now study how the ratio $\sigma_{\rm max}/\sigma_{\rm rms}$ depends on the surface roughness power spectra.
We will consider two cases where there is (a) no roll-off region in the power spectra and (b) where a roll-off region occur.
Fig. \ref{1logq.2logC.with.and.without.roll.off.H=1.eps}
shows the surface roughness power spectra as a function of the wave number
(log-log-scale) used in the calculations of the surface height profile
for surfaces with the Hurst exponent $H=1$ without (a) and with (b) a roll-off region.
Note that a vertical shift in power spectra in (b) has no influence on the ratio $\sigma_{\rm max}/\sigma_{\rm rms}$
since it correspond to scaling $C({\bf q})$ with some factor $s^2$, 
which is equivalent to scaling $h({\bf x})$ and hence  $\sigma ({\bf x})$ with the factor of $s$,
which changes both $\sigma_{\rm max}$ and $\sigma_{\rm rms}$ with the same 
factor $s$, so the ratio $\sigma_{\rm max}/\sigma_{\rm rms}$ is unchanged.

\begin{figure}
        \includegraphics[width=0.45\textwidth]{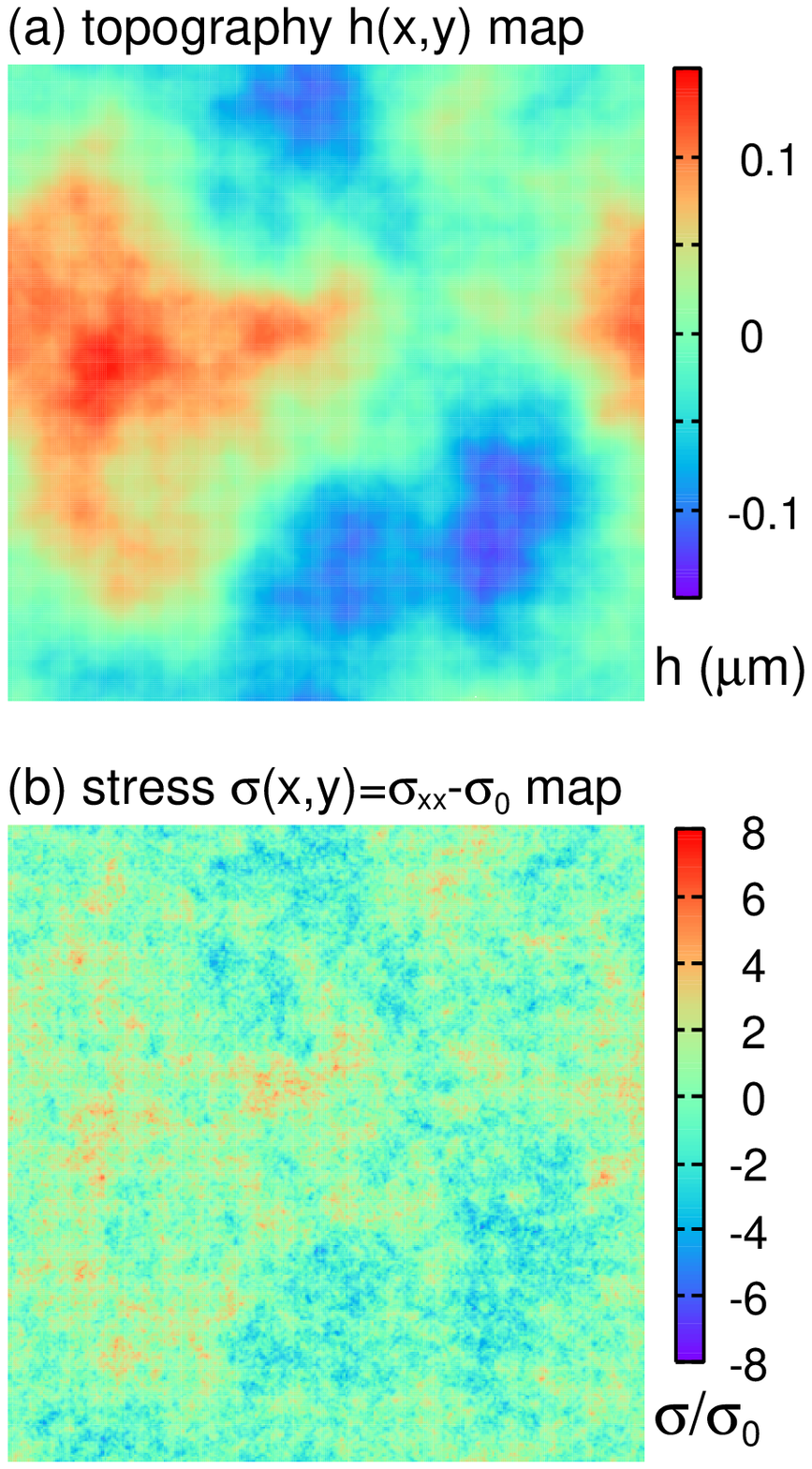}
        \caption{\label{heightmap.6.norolloff.eps}
The (a) height topography $z=h(x,y)$ (where $h$ positive into the material) and (b) the surface stress
distribution $\sigma (x,y) = \sigma_{xx}(x,y)-\sigma_0$ for the surface with 
$L=1.26 \ {\rm \mu m}$ without a roll-off.
The rms surface roughness 
$h_{\rm rms} = 0.058 \ {\rm \mu m}$, the rms slope $\xi = 0.823$, 
and the rms stress 
$\sigma_{\rm rms} = 1.646 \sigma_0$.
Note on the average stress tend to be highest in the deep roughness wells (red area in both
pictures).
}
\end{figure}

\begin{figure}
        \includegraphics[width=0.45\textwidth]{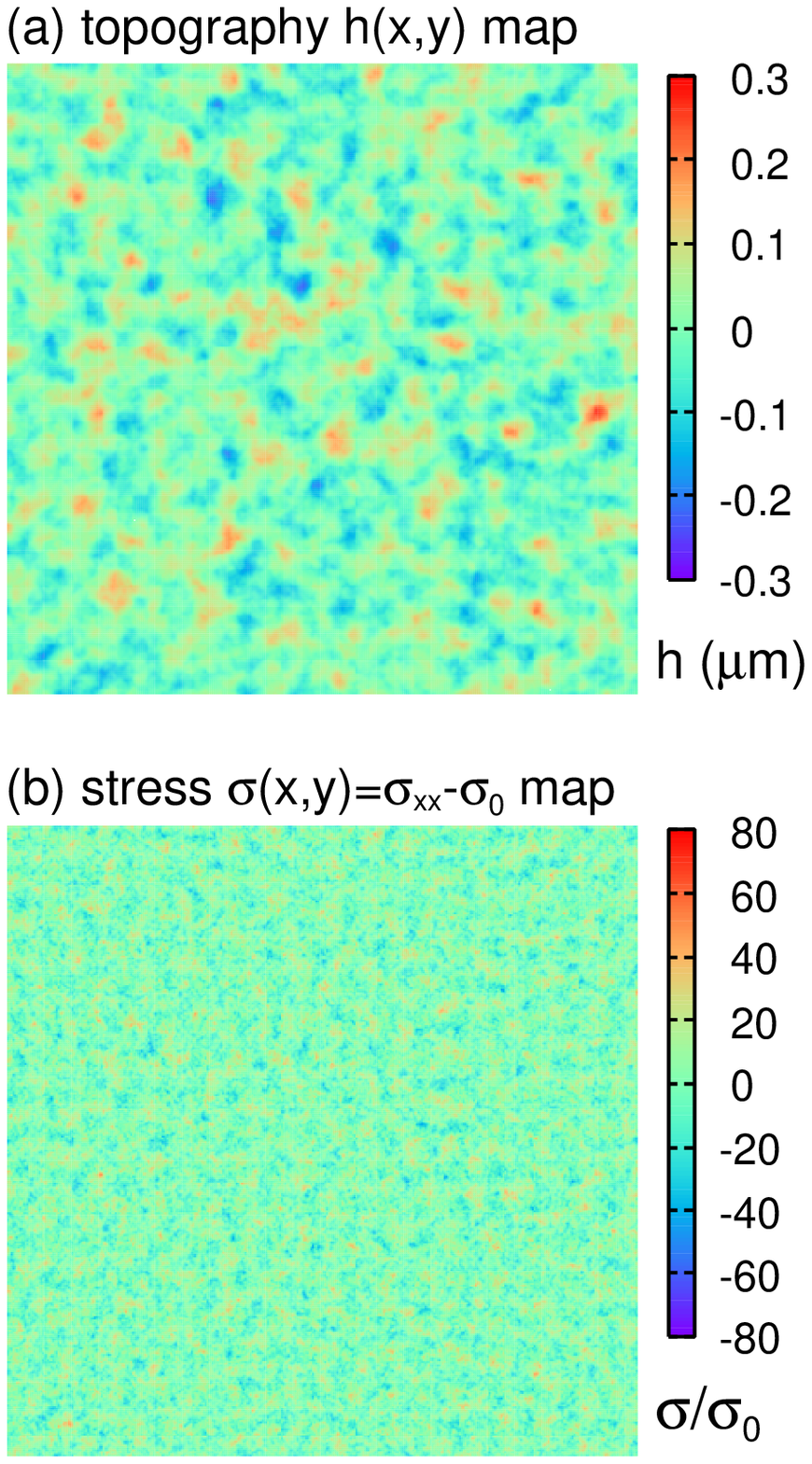}
        \caption{\label{plot.stress.roll.off.5.eps}
The same as in Fig. \ref{heightmap.6.norolloff.eps} but with roll-off.
The (a) height topography $z=h(x,y)$ (where $h$ positive into the material) and (b) the surface stress
distribution $\sigma (x,y) = \sigma_{xx}(x,y)-\sigma_0$ for the surface with 
$L=1.26 \ {\rm \mu m}$.
The rms surface roughness 
$h_{\rm rms} = 0.058 \ {\rm \mu m}$ and the rms slope $\xi = 10.86$.
The large slope (and hence large $\sigma/\sigma_0$) 
is unphysical and result from the fact that the rms roughness amplitude was chosen the same
for the power spectrum with and without roll-off. However, scaling $h(x,y)$ by a factor of 
$0.1$ gives a physical reasonable slope and this correspond to scaling the stress with the
same factor of 0.1 which would give a similar stress variation as in the case of no roll-off. 
}
\end{figure}

Fig. \ref{1ratio.Pmax.over.prms.NoRollOff.eps}
shows the cumulative probability for the 
ratio $\sigma_{\rm max}/\sigma_{\rm rms}$ between the height of the highest asperity (relative to the average surface plane)
and the rms roughness amplitude for the power spectra shown in Fig. \ref{1logq.2logC.with.and.without.roll.off.H=1.eps} 
without (a) and with (b) a roll-off region. Note that $\sigma_{\rm max}/\sigma_{\rm rms}$ depends on the system size in a very similar way
for the case of no roll-off region and a roll-off region. This is very different from the roughness amplitude 
ratio $h_{\rm max}/h_{\rm rms}$ which is independent of the size of the surface area when no roll-off occur.
For the case of a roll-off region the ratio $h_{\rm max}/h_{\rm rms}$ increases continuously with increasing roll-off 
region $q_0 < q < q_{\rm r}$ as also observed for $\sigma_{\rm max}/\sigma_{\rm rms}$.
        
Fig. \ref{1logLx.2ratio.sigMax.sigRMS.with.theory.eps}
shows the ratio $\sigma_{\rm max}/\sigma_{\rm rms}$ between the highest surface stress and the rms surface stress.
as a function of the logarithm of the size of the unit $L$. The red and green lines
are with and without a roll-off region in the power spectra, and the results are obtained after averaging over 60 realizations of the
surface roughness. The blue line is the theory prediction using (15) with $N=(L/a)^2$ i.e. assuming that all the roughness components
contribute equally. 

We note that since $\sigma_{\rm rms}$ is an average over the whole surface area it is nearly identical for all the 60 realizations.
This is clear if we plot the probability distribution of stresses as shown in 
Fig. \ref{1sigma.over.sigma0.2Probability.eps}. All 60 realizations gives nearly
perfect Gaussian distributions with equal width (the rms width is $\sigma_{\rm rms}$).

Fig. \ref{heightmap.6.norolloff.eps}
shows the (a) height topography 
$z=h(x,y)$ (where $h$ positive into the material) and (b) the surface stress
distribution $\sigma (x,y) = \sigma_{xx}(x,y)-\sigma_0$ for the surface with 
$L=1.26 \ {\rm \mu m}$ without a roll-off.
The rms surface roughness 
$h_{\rm rms} = 0.058$, the rms slope $\xi = 0.823$, 
and the rms stress 
$\sigma_{\rm rms} = 1.646 \sigma_0$.
Note on the average stress is high in the deep roughness wells (red area in both
pictures).
        
Fig. \ref{plot.stress.roll.off.5.eps}
shows the same as in Fig. \ref{heightmap.6.norolloff.eps} but with roll-off.
The rms surface roughness 
$h_{\rm rms} = 0.058$ and the rms slope $\xi = 10.86$.
The large slope (and hence large $\sigma/\sigma_0$) 
is unphysical and result from the fact that the rms roughness amplitude was chosen the same
for the power spectrum with and without roll-off. However, scaling $h(x,y)$ by a factor of 
$0.1$ gives a physical reasonable slope and this correspond to scaling the stress with the
same factor of 0.1 which would give a similar stress variation as in the case of no roll-off. 

Fig. \ref{heightmap.6.norolloff.eps} shows that the highest stress surface regions tend to occur at the
bottom of the longest wavelength (large amplitude) roughness components. This support the empirical attempts to
relate the stress concentration factor to maximum height parameters such as $R_z$. However, the treatment in
Sec. 5 shows that the important parameter is the rms-slope and the range of roughness components which determines
$N$ both of which can be obtained from the surface roughness power spectra.

\vskip 0.3cm
{\bf 7 Discussion}

We have assumed that only elastic deformations occur in the solid. This is a good assumption as long as
the applied stress $\sigma_0$ is small enough but in general one expect plastic flow and crack formation and propagation at the surface.
Assume first that only plastic deformations occur (with the yield stress in tension $\sigma_{\rm Y}$).
If the applied stress is larger than $\sim \sigma_{\rm Y} /10$ the local stress in some locations will be 
above the plastic yield stress in tension and plastic deformation is expected. 
But this plastic flow may occur only at short length scale (involving the short wavelength roughness), 
and since the yield stress may increase at short length scale the plastic deformations may be 
smaller than expected based on the macroscopic yield stress. 
(An increase in the yield stress at short length scales is well-known from indentation experiments where
the penetration hardness $\sigma_{\rm P} \approx 3 \sigma_{\rm Y}$ often increases as the indentation size 
decreases.\cite{size})
If plastic deformation occur it will change the surface profile and reduce the local tensile 
stress so that it is at most the yield stress in tension $\sigma_{\rm Y}$. 
If one assume that the plastically deformed surface region 
does not change the stress in the regions which have not undergone
plastic deformation, and if $\sigma_{\rm Y}$ is independent of the length scale, then the fraction of the
surface which has undergone plastic deformation is determined by
$${A_{\rm pl} \over A_0} = \int_{\sigma_{\rm Y}-\sigma_0}^\infty d\sigma \ P(\sigma ) $$

\begin{figure}[!ht]
        \includegraphics[width=0.48\textwidth]{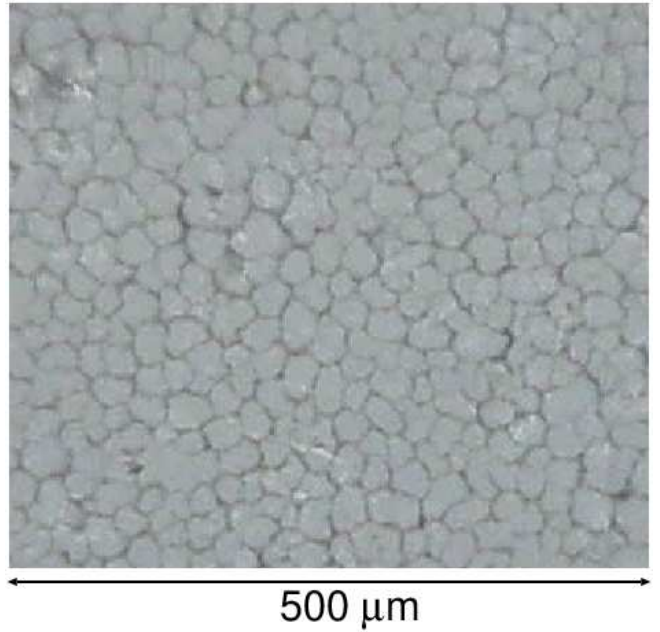}
\caption{\label{crack}
Optical picture of the sandblasted glass surface after annealing at $860^\circ {\rm C}$ for 1 hour.
Note the cell-like structure of the surface which we interpret as a network of short cracks. 
For a smooth glass plate the same annealing cycle 
result in a very low concentration of cracks. From \cite{Si}. 
}
\end{figure}

Next we consider the ideal case where no plastic deformations occur but only crack propagation (ideal brittle solid).
The stress concentration at the surface resulting from the surface roughness can
initiate crack growth. Here it is important to note that 
the deformation (and stress) field from a roughness component with the wavelength $\lambda$ will extend into the solid
a distance $\sim \lambda /\pi$ [see (C45) and Appendix E] so when the crack has moved into the solid 
the distance $\sim \lambda /\pi$ it has already made use of all the
elastic deformation energy associated with this roughness component. 
If it can propagate further depends on the elastic energy stored in 
the other longer roughness wavelength components. Hence in this case it is possible that a crack propagate only a finite 
distance into the solid. This could result in a network of surface cracks of finite depth as sometimes observed
in experiments. Thus in Ref. \cite{Si} it was found that when a sandblasted silica glass plate was thermally annealed
a network of short cracks formed on the surface (see Fig. \ref{crack}). 
This could be due to a tensile stress acting in a top layer of the
silica plate due to the thermal contraction during cooling, which is stronger at the
(colder) surface than inside the glass plate. For a similar glass plate 
which was not sandblasted a much smaller concentration
of cracks was formed, as expected from theory due to a lower concentration of surface defects.

\begin{figure}[!ht]
        \includegraphics[width=0.25\textwidth]{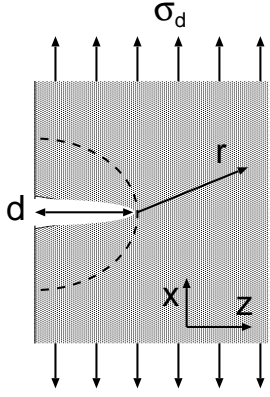}
\caption{\label{SurfaceCrackDiscuss.eps}
A crack of length $d$ at the surface of an elastic solid. The stress a distance $\sim d$ from the crack
tip is of order $\sigma_d$. The crack reduces the elastic energy density to nearly zero
within a volume element (dashed line) with volume $\sim d^2 L$, where $L$ is the length of the crack in the
$y$-direction.
}
\end{figure}

\begin{figure}[!ht]
        \includegraphics[width=0.33\textwidth]{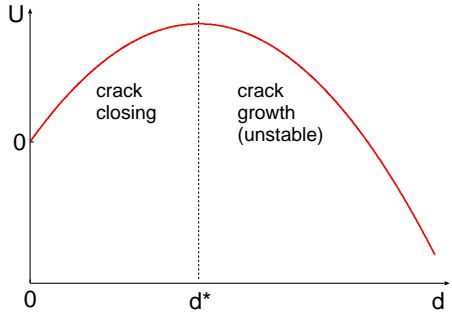}
\caption{\label{1d.2U.crack.eps}
The total energy $U(d)$ as a function of the crack length $d$.
For $d>d^*$ the total energy decreases with increasing $d$ resulting in
unstable (accelerating) crack growth.
}
\end{figure}

Here we present a simple dimensional analysis of how the surface stress generate cracks and plastic
deformations. Consider a small crack at the surface of a solid exposed to a tensile stress. Except for an angular factor,
of no importance here, the stress in the vicinity of the crack tip is\cite{crack}
$$\sigma \approx \sigma_d \left ( {d\over r} \right )^{1/2}\eqno(19)$$
where $r$ is the distance from the crack tip,
$d$ is the length of the crack and $\sigma_d$ is the tensile stress a distance $\sim d$
from the crack tip (note: $\sigma_d$ is larger than the applied stress $\sigma_0$ which occur far away from the crack tip). 
The critical length of the crack is determined by standard arguments\cite{Griffith}, namely $U'(d)=0$ where $U(d)$ is the
total energy. The reduction in the elastic energy induced by the crack
$$U_{\rm el} \approx - {1\over 2} \sigma \epsilon d^2 L  \approx  - {1\over 2} {\sigma_d^2 \over E} d^2 L$$
where $d^2L$ is the volume where the deformation energy is reduced (see Fig. \ref{SurfaceCrackDiscuss.eps}). The surface energy
$$U_{\rm area} = dL \gamma$$
where $\gamma$ is the energy per unit area to create the fracture surfaces.
From $U'(d)=0$ with $U=U_{\rm el}+U_{\rm area}$ we get the critical length $d=d^*$
$$d^* = {E\gamma \over \sigma_d^2}\eqno(20)$$
Fig. \ref{1d.2U.crack.eps} shows the total energy $U(d)$ as a function of $d$.
If $d<d^*$ no crack growth will occur while when $d>d^*$ unstable (accelerating) crack growth may occur.
However, since the tensile stress decreases with increasing distance into the solid the crack will
propagate only as long as the drop in the elastic energy is larger than the increase in the surface energy.
We will now study this using a simple model.

\begin{figure}[!ht]
        \includegraphics[width=0.48\textwidth]{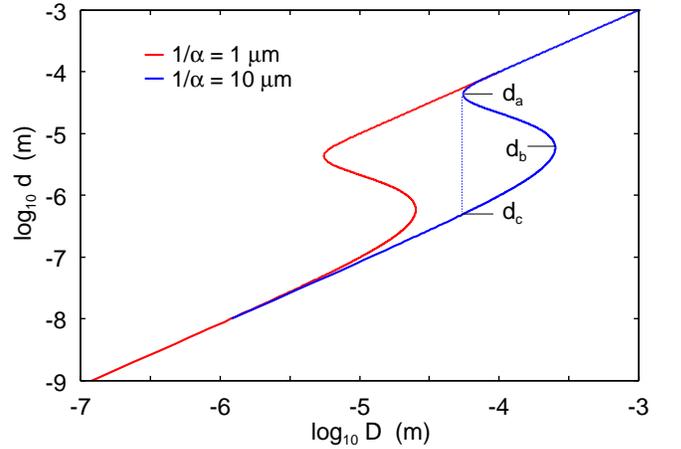}
\caption{\label{1logd.2logd1.eps}
Solution to (21) where $D=E \gamma/\sigma_0^2$. For $\beta = 10$ and $1/\alpha = 1 \ {\rm \mu m}$ (red line)
and $10  \ {\rm \mu m}$ (blue line).
}
\end{figure}

The stress at the surface decay with the distance $z$ into the solid.
As an example, assume that the stress $\sigma (z)$ decreases from $(1+\beta)\sigma_0$ to $\sigma_0$ with the distance
$z$ according to
$$\sigma (z) = \left (1+\beta e^{-\alpha z} \right ) \sigma_0$$ 
In this case (20) gives
$$d \left (1+\beta e^{-\alpha d} \right )^2 =  D\eqno(21)$$
where the length parameter $D=E\gamma /\sigma_0^2$.
In Fig. \ref{1logd.2logd1.eps} we show 
the solution to (21) for $\beta=10$ and $1/\alpha = 1 \ {\rm \mu m}$ (red curve) and $10 \ {\rm \mu m}$ (blue curve).

Suppose now that we slowly increase the external stress $\sigma_{xx}=\sigma_0$ until a crack-like defect (with the initial 
length $d_1$) start to grow. At this point we keep $\sigma_0$ fixed and study the time evolution of the
crack length $d$. As $\sigma_0$ increases $D$ decreases from $\infty$ to some finite value $D$. 
Assume first  $d_1 > d_{\rm a}$. The crack cannot grow until we increased
$\sigma_0$ so that  $d(D) = d_1$. At this point the elastic energy stored in the vicinity of the crack tip
is big enough to break the bonds and allow the crack to grow. However, as it growth ($d$ increases) Fig.
\ref{1logd.2logd1.eps} shows that a larger $D$, and hence smaller applied stress $\sigma_0$, 
is enough to grow the crack further. But since we keep
$\sigma_0$ (and hence $D$) fixed the crack will accelerate
resulting in a rapid catastrophic fracture of the solid. The same is true if the initial crack length is $d_1 < d_{\rm c}$

Now assume $d_{\rm b} < d_1 < d_{\rm a}$.
The crack does not grow until $D$ has decreased so that $d(D) = d_1$. At this point the crack start to grow
but now an increase in the crack length require a smaller $D$, and hence larger $\sigma_0$, i.e. there is not enough
stored elastic energy to propagate the crack if $\sigma_0$ is kept constant.
If we increase $\sigma_0$ the crack will grow but in a stable manner until the crack length reach $d=d_{\rm a}$ 
at which point fast (accelerated) growth occur again resulting in catastrophic failure of the body.

Finally, assume that $d_{\rm c} < d_1 < d_{\rm b}$. In this case when $D$ has decreased (and the stress $\sigma_0$ has increased) 
so that $d(D)=d_1$ the crack length will increase initially in an accelerating way since the $d(D)$ curve has a positive slope
at $d=d_1$. However, since $D > D_{\rm a}$, where $D_{\rm a}$ is the solution to $d(D)=d_{\rm a}$ the motion will slow down and stop
somewhere in the region $d_{\rm b} < d_1 < d_{\rm a}$. 
Here we have neglected kinetic effects i.e. we have assumed that there is
not enough kinetic energy associated with the initial rapid crack tip
motion to move over the ``barrier'' at $d=d_{\rm a}$.
(Note: Linear elastic fracture mechanic theory predict that cracks have no inertia\cite{inert}. 
Thus the crack will adjust its speed instantaneously to the driving force determined by the elastic energy stored in the solid
in its vicinity. If the elastic deformation energy driving crack propagation is larger than
the adiabatic fracture energy $\gamma$ then the additional energy is ``dissipated'' 
by creating surface roughness (and hence surface area) on the fracture surfaces, 
and by emission of elastic waves from the crack tip, and by other inelastic processes. 
However, see Ref. \cite{inertia}.)
 
The discussions above assumes that no plastic deformations occur during crack propagation. For most solids, in particular metals,
some plastic deformation (or other inelastic processes) will occur close to the crack tip\cite{plast,Irwin}. One can determine
the size $d_{\rm Y}$ of the region where plastic flow occur as follows: Plastic flow start when the tensile stress
reaches $\sigma_{\rm Y}$. Using (19) we get
$$\sigma_{\rm Y} \approx \sigma_d \left ( {d\over d_{\rm Y}} \right )^{1/2}$$
or using (20)
$$d_{\rm Y} = \left ({\sigma_d \over \sigma_{\rm Y}} \right )^2 d = {E \gamma \over \sigma_{\rm Y}^2}\eqno(22)$$
If $d_{\rm Y} << d$ then the crack theory presented above is valid but the surface energy $\gamma$ is 
not just the energy to break the bonds at the crack tip but must include the energy of 
plastic deformation (the crack surfaces are covered by thin films of plastically deformed material).
If $d_{\rm Y} > d$ no crack propagation will occur but just local plastic deformation. 
For amorphous solids such as silica glass and amorphous silicon $d_{\rm Y}$ is typically a few ${\rm nm}$ while for metals
$d_{\rm Y} \approx 10 \ {\rm \mu m}$ or more (see Appendix F). 

Similar ideas as discussed above have been presented in models 
of adhesive wear where big wear particles form by crack propagation 
in the large asperity contact regions, 
while small asperity contact regions deform plastically without generation of
wear particles\cite{Rabin0,Rabin1,Rabin2,Moli2,Moli0,Moli1}.

The stress concentration due to surface roughness can result in stress corrosion\cite{Cic}.
Chemical bonds between atoms can be broken either by thermal fluctuations or by an applied
force (stress). When the applied force is not high enough to break a bond the bond could still be 
broken by a large enough thermal fluctuation\cite{Szl}. 
When the applied force increases the energy needed to
overcome the barrier towards bond breaking decreases and the probability rate of (thermally assisted) bond 
breaking increases. This stress-aided, thermally activated process can result in the slow growth 
of surface cracks and to stress corrosion.

Stress corrosion cracking is the formation of cracks in a 
material through the simultaneous action of a tensile stress, temperature and a corrosive environment.
Stress corrosion cracking has become one of the main reasons for the failure of steam generator tubing.
The specific environment is of crucial importance, and only very 
small concentrations of certain highly active chemicals are 
needed to produce catastrophic cracking, often leading to devastating and unexpected failure.

Finally we note that the main driving force for the study of surface stress concentration is material fatigue
which account for the majority of disastrous failure of mechanical devices e.g. airplanes. 
Fatigue damage of a component typically develop due to surface stress concentration originating
from the surface topography\cite{fat1,fat2}. This result in the formation of crack-like defects which at some stage
can propagate rapidly, possibly resulting in an unexpected catastrophic event.

Fatigue crack propagation in metals involves stress concentration and plastic deformations. 
Short wavelength roughness may be ``smoothed'' by plastic flow before a crack can nucleate and propagate because
the elastic deformation energy density needed to propagate a crack increases as the crack size decreases.

It is remarkable that a solid can fail by crack propagation when exposed to a stress fluctuating in time
(fatigue failure), but not (if the stress is small enough) when
exposed to a static stress of the same magnitude as the amplitude of the oscillating stress. This indicate that
some irreversible processes, not involving crack propagation, 
occur during the stress oscillations. For metals this likely involves point defects and
dislocations which can form and move by the oscillating crack tip stress field, and which accumulate with increasing time
in the region close to the crack tip and reduce the energy per unit area $\gamma$ to create new fracture surfaces\cite{fatigue}.
If $\gamma$ is reduced enough the crack can propagate even if for the original virgin solid this was not the case.
For viscoelastic materials such as rubber the effective energy $\gamma$ to propagate a crack is smaller in an oscillating stress field
because of viscoelasticity\cite{visc1,visc2}, and this explain why rubber wear, involving removing small rubber particles, occur during
sliding (where the rubber surface is exposed to pulsating stresses from the countersurface asperities) while for a static
contact with the same stress amplitude no (or negligible) crack propagation and wear particle formation occur. 

Other applications of the theory presented above are to surface kinetics\cite{d3}. The atoms in a stressed region on a solid surface
have higher energy than in a non-stressed region. As a result less energy is needed to remove atoms from stressed
surface regions. This may result in diffusion of atoms from stressed regions to less stressed surface regions.
For a flat surface the surface stress is uniform (equal to $\sigma_0$) but for a surface with roughness the stress varies with
the surface position, and theory shows that this may result in short wavelength roughness being smoothed by surface diffusion while long
wavelength roughness may grow unstably. Similarly, evaporation-condensation is affected by the surface stress. Thus when the
surface evolution is controlled by evaporation from or condensation to a surface, such that there is no
net translation of the surface, the short wavelength roughness are smoothed by the evaporation/condensation
process, whereas long wavelength roughness grow unstably\cite{d3}.

The theory in this paper is based on the small slope approximation. 
In Ref. \cite{d1} the results of the small slope
approximation was compared to experiment and to FEM calculations 
for 1D wavy surfaces, and nearly perfect agreement with the
theory was obtained for surfaces with the rms-slope $\sim 0.2$, 
where the maximum stress concentration factor was $S\approx 2$.
Similarly, in Ref. \cite{add} the theory prediction
was found to be within $\sim 20\%$ of the FEM prediction even for a 1D wavy surface
with the rms slope as large as $\sim 1$.

\vskip 0.3cm
{\bf 8 Summary and conclusion}

When a body is exposed to external forces large local stresses may occur at the surface because of surface
roughness. For randomly rough surfaces I calculate the probability distribution of surface stress in response to a
uniform external tensile stress $\sigma_0$. 
I have shown that for randomly rough surfaces of elastic solids,
the maximum local surface stress is given by $(1+ s \xi)\sigma_0$, 
where typically $s\approx 10$. For most surfaces of engineering interest,
when including all the surface roughness, the rms slope $\xi \approx 1$ giving 
maximal local tensile stresses of order $\sim 10 \sigma_0$ or more.

I have presented numerical simulation results for the stress distribution $\sigma (x,y)$
and discussed the role of the stress concentration on plastic deformation and surface crack 
generation and propagation. The present study is important for many application and in
particular for fatigue due to pulsating external forces, and to surface kinetics such as
surface diffusion and evaporation/condensation phenomena.

\vskip 0.3cm
{\bf Acknowledgments:}
I thank Jay Fineberg for discussions about crack inertia and R.O. Jones and M. M\"user for discussions
about chemical bonding in relation to Appendix F. I thank R. Carpick for comments on the text.

\vskip 0.1cm
{\bf Funding:} Open Access funding enabled and organized by Projekt
DEAL. The authors have not disclosed any funding.

\vskip 0.1cm
{\bf Conflict of interest:}
The author declare no conflict of interest in this study.

\vskip 0.3cm
{\bf Appendix A: The power spectra}

In Ref. \cite{PT} (see also \cite{P,PJCP}) we have derived (4) but for the readers convenience
we repeat the derivation here. Because of translation invariance of the statistical properties of a 
randomly rough surface we can write (1) as
$$C({\bf q})  = {1\over (2\pi )^2} \int d^2 x \ \langle h({\bf x}+{\bf x}') h({\bf x}')\rangle e^{i{\bf q}\cdot {\bf x}}\eqno(A1)$$
Since (A1) is independent of ${\bf x}'$ we can integrate over the ${\bf x}'$-surface and divide by the nominal area $A_0$ to get
$$C({\bf q})  = {1\over (2\pi )^2} {1\over A_0} \int d^2 x d^2 x' \ \langle h({\bf x}+{\bf x}') 
h({\bf x}')\rangle e^{i{\bf q}\cdot {\bf x}}$$
Using (3) and performing the ${\bf x}$ and ${\bf x}'$ integrals and using that
$${1\over (2\pi )^2} \int d^2x \ e^{i({\bf q}+{\bf q'})\cdot {\bf x}}=\delta ({\bf q}+{\bf q'})$$
we get
$$C({\bf q})  = {(2\pi )^2\over A_0} \int d^2q' d^2q'' \ \langle h({\bf q'}) h({\bf q''}) \rangle $$
$$\times \delta ({\bf q'}+{\bf q}) \delta ({\bf q'}+{\bf q''})$$
$$= {(2\pi )^2 \over A_0} \langle h({\bf q}) h(-{\bf q})\rangle $$

\begin{figure}
        \includegraphics[width=0.3\textwidth]{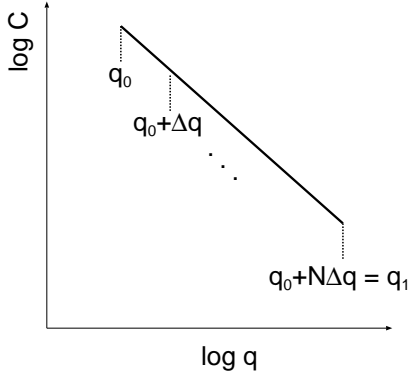}
        \caption{\label{devide.eps}
The surface roughness power spectrum is divided into $N$ segments covering all
length scales.
}
\end{figure}

\vskip 0.3cm
{\bf Appendix B: Stress concentration factor}

The stress at the tip of a surface ``cavity'' (or valley) is larger than the applied stress
$\sigma_0$ by a factor $S=1+2\surd (d_0/r_0)$ (see Fig. \ref{cavity.eps}). But real surface have roughness
on many length scales which we can formally consider as the sum of
$N$ wavenumber regions as indicated in Fig. \ref{devide.eps}. If $S_1$ is the enhancement
factor including only the roughness from the longest wavelength segment $q_0 < q <q_0+\Delta q$ then
when we add the roughness from the next roughness segment the enhancement becomes
$S_1 S_2$ and so on. If the segments are short then for each length scale $\surd (d/r) << 1$
and we get the total stress enhancement factor
$$S =(1+2\surd (d/r)\mid_1 ) (1+2\surd (d/r)\mid_2 )..(1+2\surd (d/r)\mid_N ) $$
$$\approx 1+\sum_n 2\surd (d/r)\mid_n$$
The qualitative picture underlying this approach is similar to the way multiscale roughness
was taken into account in the study of fluid contact angles on randomly rough surfaces in Ref.
\cite{fangle}.

\begin{figure}
        \includegraphics[width=0.27\textwidth]{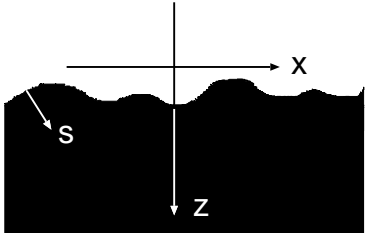}
        \caption{\label{coordinate.eps}
The vector ${\bf s}$ is normal to the surface $z=h(x,y)$ and point into the material.
The $z$-axis is normal to the average surface plane with the positive axis into the material.
}
\end{figure}

\vskip 0.3cm
{\bf Appendix C: Stress--surface-roughness relation}

Here we derive the relation between the stress $\sigma_{xx}({\bf x})=\sigma_0+\sigma ({\bf x})$ 
and the surface roughness $z=h({\bf x})$. This problem has been studied before for a 1D roughness
profile using the Airy stress function\cite{d1} (see also \cite{d2} for another approach),
but here we derive it for an arbitrary 2D surface roughness 
profile in the small slope approximation. The derivation presented here can be 
easily generalized to layered materials\cite{layer}.

To first order in $h'_x$ and $h'_y$ the normal unit vector to the surface $z=h(x,y)$
is given by (see Fig. \ref{coordinate.eps}): 
$${\bf s}=\left (-h'_x,-h'_y,1 \right ).$$
Since the surface stress $\sigma_{ij} s_j$ must vanish we get 
$$-h_x' \sigma_{xx} - h'_y \sigma_{xy}+\sigma_{xz}=0$$
$$-h_x' \sigma_{yx} - h'_y \sigma_{yy}+\sigma_{yz}=0$$
$$-h_x' \sigma_{zx} - h'_y \sigma_{zy}+\sigma_{zz}=0$$
Assuming that without the surface roughness the stress $\sigma_{xx}= \sigma^0_{xx}$ 
and $\sigma_{yy}= \sigma^0_{yy}$ are constant while all the other stress components vanish.
This imply that with surface roughness all other stress components are already of 
first order in $h_x'$ and $h_y'$, and products such as
$h_x' \sigma_{xy}$ are of second order and can be neglected. Thus 
to first order in $h'_x$ and $h'_y$ 
$$-h'_x \sigma^0_{xx} + \sigma_{xz} =0 \eqno(C1)$$
$$-h'_y \sigma^0_{yy} + \sigma_{yz} = 0 \eqno(C2)$$
$$\sigma_{zz} =0 \eqno(C3)$$
In what follows we will denote $\sigma_{xx}-\sigma^0_{xx}$ with just $\sigma_{xx}$ and similar for 
$\sigma_{yy}$. In this case all the components of the stress tensor $\sigma_{ij}$ will be of first order in $h'_x$ and $h'_y$.
Since the stress tensor is already linear in $h_x'$ and $h_y'$ we can consider the surface of the solid
as flat (no roughness) when calculating the elastic deformation field and the stress in the solid using the
boundary conditions (C1)-(C3).

We write
$$h({\bf x})= \int d^2 q \ h({\bf q}) e^{i {\bf q}\cdot {\bf x}}\eqno(C4)$$
so from (C1) the stress $\sigma_{xz} ({\bf x},z)$ at the surface $z=0$ takes the form
$$\sigma_{xz} ({\bf x},0)= \sigma^0_{xx} \int d^2 q \ (iq_x) h({\bf q}) e^{i {\bf q}\cdot {\bf x}}\eqno(C5)$$
and similar for $\sigma_{yz} ({\bf x},0)$.
If we define the vector ${\pmb \sigma} = (\sigma_{xz},\sigma_{yz},\sigma_{zz})$
the boundary conditions (C1)-(C3) can be written as
$${\pmb \sigma}  = (h'_x \sigma^0_{xx},h'_y \sigma^0_{yy},0) \eqno(C6)$$
for $z=0$. 

To calculate $\sigma_{xx}$ to first order in $h'_x$ and $h'_y$ we must solve the equations of elasticity for a semi-infinite solid
with the stress ${\pmb \sigma}$ acting on the surface $z=0$. We choose a coordinate system $xyz$ with $z=0$ in the surface plane
and the positive $z$-axis pointing into the solid. Let ${\bf n}$ be a unit vector along the $z$-axis.
Following Ref. \cite{Persson2} we write the displacement field as
$${\bf u} = {\bf p} A+ {\bf K} B +{\bf p}\times {\bf K} C\eqno(C7)$$
where $A$, $B$ and $C$ are three scalar fields and where ${\bf p} = - i \nabla$, ${\bf K} = {\bf n}\times {\bf p}$ and 
${\bf p}\times {\bf K}$ are three ``orthogonal'' vector operators. For mathematical convenience we will assume that
$h({\bf x})$ varies slowly in time as ${\rm exp} (-i\omega t)$ and we will take the $\omega \rightarrow 0$ limit
at the end of the calculation. The advantage of this approach is that we do not need to use a biharmonic-type of
equation for the displacement field but rather the simpler wave equations (see Ref. \cite{Persson2}):
$$\left (\omega^2 +c_L^2\nabla^2 \right )A = 0\eqno(C8)$$
$$\left (\omega^2 +c_T^2\nabla^2 \right )B = 0\eqno(C9)$$
$$\left (\omega^2 +c_T^2\nabla^2 \right )C = 0\eqno(C10)$$
with the general solutions
$$A({\bf x},z,t) = \int d^2q d\omega \ A({\bf q},\omega) e^{i({\bf q}\cdot {\bf x}+p_L z -i\omega t)}\eqno(C11)$$
$$B({\bf x},z,t) = \int d^2q d\omega \ B({\bf q},\omega) e^{i({\bf q}\cdot {\bf x}+p_T z -i\omega t)}\eqno(C12)$$
$$C({\bf x},z,t) = \int d^2q d\omega \ C({\bf q},\omega) e^{i({\bf q}\cdot {\bf x}+p_T z -i\omega t)}\eqno(C13)$$
where
$$p_T=\left ({\omega^2\over c_T^2} - q^2 \right )^{1/2}, \ \ \ \ \ p_L=\left ({\omega^2\over c_L^2} - q^2 \right )^{1/2}\eqno(C14)$$
In what follows for simplicity we will  suppress the frequency argument and write $A({\bf q})$ instead of $A({\bf q},\omega)$,
and similar for other quantities.
The transverse and the longitudinal sound velocities, $c_T$ and $c_L$, can be related to the Lame elasticity parameters
$\mu$ and $\lambda$ as
$${c_L^2 \over c_T^2} = {\lambda \over \mu}+2, \ \ \ \ \ {\mu \over \lambda} = {1-2\nu \over 2 \nu}\eqno(C15)$$
where $\nu$ is the Poisson ratio. Using these equations one get
$$ {\lambda \over \mu} \left ({c_L^2\over c_T^2}-1\right )^{-1} = 2\nu, \ \ \ \ \ 
\left (1-{c_T^2\over c_L^2} \right )^{-1} = 2 (1-\nu) \eqno(C16)$$

We consider first the case when the rectangular block is elongated in the $x$-direction with
$\sigma^0_{xx}=\sigma_{0}$. 
In this case
$${\pmb \sigma} ({\bf x}) = (h'_x ,0,0) \sigma_0$$
and 
$${\pmb \sigma} ({\bf q}) = i {\bf e}_x q_x h({\bf q})\sigma_0\eqno(C17)$$
where ${\bf e}_x$ is a unit vector along the $x$-axis.
Substituting this in (A18)-(A20) in Ref. \cite{Persson2} gives
$$A({\bf q}) ={1\over \mu S} 2p_T q_x^2 h({\bf q})\sigma_0 \eqno(C18)$$
$$B({\bf q}) =- {1\over \mu} {q_x q_y \over q^2 p_T} h({\bf q})\sigma_0 \eqno(C19)$$
$$C({\bf q}) = -{1\over \mu S} \left ({\omega^2\over c_T^2}-2q^2 \right ) {q_x^2\over q^2} h({\bf q}) \sigma_0\eqno(C20)$$
where
$$S = \left ({\omega^2\over c_T^2}-2q^2\right )^2 +4 q^2p_Tp_L\eqno(C21)$$
Using that as $\omega \rightarrow 0$ to leading order in $\omega$
$$p_T = i q \left (1- {\omega^2 \over c_T^2 q^2}\right )^{1/2} \approx iq - iq {1\over 2} {\omega^2 \over c_T^2 q^2}\eqno(C22)$$
and similar for $p_L$ we get as $\omega \rightarrow 0$
$$S \approx 2 q^2 \omega^2 \left ({1\over c_L^2}- {1\over c_T^2} \right )\eqno(C23)$$

The stress tensor
$$\sigma_{ij} = \mu \left (u_{i,j}+u_{j,i} \right ) +\lambda u_{k,k} \delta_{ij}\eqno(C24)$$
We are interested in the $\sigma_{xx}$ stress component which can be written as
$$-i\sigma_{xx}= 2\mu p_x u_x +\lambda {\bf p} \cdot {\bf u}\eqno(C25)$$
or using (C7) we get 
$$-i\sigma_{xx} ({\bf x},z) = 2\mu \left (p_x^2 A -p_xp_y B -p_x^2p_z C\right ) +\lambda p^2 A$$
Using $p^2A= (\omega /c_L)^2 A$ we get for $z=0$ 
$$-i\sigma_{xx} ({\bf q},0)= 2\mu \left (q_x^2 A({\bf q})-q_xq_y B({\bf q})- p_Tq_x^2 C({\bf q}) \right )$$ 
$$+\lambda \left ({\omega \over c_L}\right )^2 A({\bf q})\eqno(C26)$$
Substituting (C18)-(C20) in this equation gives
$$-i\sigma_{xx}= {2\omega^2 \over S} p_T q_x^2 \left ({\lambda \over \mu c_L^2}+{1\over c_T^2} 
{q_x^2\over q^2} \right ) h({\bf q})\sigma_0\eqno(C27)$$
$$ +2{q_x^2 q_y^2 \over q^2 p_T}  h({\bf q})\sigma_0 $$
Using (C23) this equation gives as $\omega \rightarrow 0$ 
$$-i\sigma_{xx}=  p_T {q_x^2 \over q^2}  \left ({1\over c_L^2}- {1\over c_T^2} \right )^{-1} 
\left ({\lambda \over \mu c_L^2}+{1\over c_T^2} 
{q_x^2\over q^2} \right ) h({\bf q})\sigma_0$$
$$ +2{q_x^2 q_y^2 \over q^2 p_T}  h({\bf q})\sigma_0 \eqno(C28)$$
For $\omega =0$ we have $p_T = i q$ and
$$\sigma_{xx}=  {q_x^2 \over q^2}  \left ({1\over c_T^2}- 
{1\over c_L^2} \right )^{-1} \left ({\lambda \over \mu c_L^2}+{1\over c_T^2} 
{q_x^2\over q^2} \right ) q h({\bf q})\sigma_0 $$
$$ +2{q_x^2 q_y^2 \over q^4}  q h({\bf q})\sigma_0 \eqno(C29)$$
Using that (C16) we get
$$\sigma_{xx}= \left [{q_x^2 \over q^2} \left (2\nu +2(1-\nu) {q_x^2\over q^2} \right ) 
+2{q_x^2 q_y^2 \over q^4}\right ] q h({\bf q})\sigma_0 \eqno(C30)$$ 
Using that $q^2=q_x^2+q_y^2$ this gives
$$\sigma_{xx}= 2 q f({\bf q}) h({\bf q})\sigma_0 \eqno(C31)$$
where
$$f= {q_x^2 \over q^2} \left (1+ \nu {q_y^2\over q^2} \right ) 
= {\rm cos}^2\theta \left(1 +\nu {\rm sin}^2 \theta\right )$$
where $q_x=q {\rm cos}\theta$ and $q_x=q {\rm sin}\theta$ is the wavevector expressed in polar coordinates.
In a similar way one can show that
$$\sigma_{yy}= \nu \left ({q_x\over q}\right )^4 2 q h({\bf q})\sigma_0 = {\nu q_x^2 \over q^2+\nu q_y^2} \sigma_{xx} \eqno(C32)$$
Note that when $\nu =0$ then $\sigma_{yy}=0$ as expected because in that limit $\sigma_{ij}= E \epsilon_{ij}$
(where $E$ is the Young's modulus) so an applied $xx$ stress is not expected to generate a $yy$ stress
response.

If surface roughness occur only in the $x$-direction then $h({\bf q}) = h(q_x)\delta (q_y)$ so that
$q^2=q_x^2$ and
$$\sigma_{xx} ({\bf q}) = 2 q h({\bf q})\sigma_0 \eqno(C33)$$ 
$$\sigma_{yy} ({\bf q}) = \nu \sigma_{xx}  ({\bf q}) \eqno(C34)$$ 
The result (C33) is the same result as obtained in Ref. \cite{d1}. 

As another example assume
$$h({\bf x}) = h_0 {\rm cos}(q_0 x) {\rm cos}(q_0 y)\eqno(C35)$$
so that
$$h({\bf q}) = {1\over (2\pi )^2} \int d^2x \ h_0 {1\over 4} \left (e^{iq_0x}+e^{-iq_0x}\right ) $$
$$\times \left (e^{iq_0y}+e^{-iq_0y}\right ) e^{-i(q_x x+q_y y)}$$
$$=h_0 {1\over 4} \left [\delta (q_x-q_0)+\delta (q_x+q_0)\right ]  \left [\delta (q_y-q_0)+\delta (q_y+q_0)\right ]$$
and
$$\sigma_{xx} ({\bf x}) = \int d^2q \ {q_x^2 \over q^2} \left (1+ \nu {q_y^2\over q^2} \right ) 2 q\sigma_0 h({\bf q}) e^{i{\bf q}\cdot {\bf x}} $$
$$=\surd 2 \left ( 1+{\nu \over 2}\right ) q_0 \sigma_0 h({\bf x}) \eqno(C36)$$
which is similar to what was found in Ref. \cite{d2} but where the term $\nu /2$ was replaced by $\nu$.
In Ref. \cite{Bar} Barber has presented a derivation of (C36) using a very different approach and obtained the same
result as found above.
Using (C32) and (C35) we get
$$\sigma_{yy} ({\bf x}) = \int d^2q \ \nu \left ({q_x\over q}\right )^4  2 q \sigma_0 h({\bf q}) e^{i{\bf q}\cdot {\bf x}}  $$
$$={\nu \over \surd 2}  q_0 \sigma_0 h({\bf x}) = {\nu \over 2+\nu}  \sigma_{xx} ({\bf x})  \eqno(C37)$$
In a typical case $\nu \approx 0.3$ and $\sigma_{yy} \approx 0.13 \sigma_{xx}$. For rubber-like materials $\nu \approx 0.5$ and
$\sigma_{yy} \approx 0.2 \sigma_{xx}$.

The mean square stress [see (11)]:
$$\langle \sigma^2 \rangle = \int d^2q q^2 (2\sigma_0)^2 C({\bf q}) f^2 ({\bf q})\eqno(C38)$$
If we assume roughness with isotropic properties then $C({\bf q})$ depends only on $q$. In this case using polar coordinates
in the integral in (C38) result in an angular integral of the form:
$${1\over 2 \pi} \int_0^{2\pi} d\theta \ f^2 = {1\over 2 \pi} \int_0^{2\pi} d\theta \ 
{\rm cos}^4\theta \left (1+\nu {\rm sin}^2 \theta \right )^2$$ 
$$ ={1\over 2 \pi} \int_0^{2\pi} d\theta \ {\rm cos}^4\theta \left (1+\nu-\nu {\rm cos}^2 \theta \right )^2$$
$$= {1\over 8} \left (3+\nu +{3\over 16}\nu^2 \right ) = g^2 (\nu)\eqno(C39)$$
where we have used that
$$I={1\over 2 \pi} \int_0^{2\pi} d\theta \ {\rm cos}^{2n}\theta = 
\left ({1\over 2}\right )^{2n} \binom{2n}{n} = \left ({1\over 2}\right )^{2n} {(2n)! \over n! n!}$$
which gives $I=1/2$, $3/8$, $5/16$ and $35/128$ for $n=1$, $2$, $3$ and $4$, respectively.
Thus we get
$$\langle \sigma^2 \rangle = \int d^2q q^2 (2\sigma_0)^2 C({\bf q}) g^2$$
$$= (2\sigma_0)^2 \xi^2 g^2\eqno(C40)$$
and the rms stress
$$\sigma_{\rm rms} = 2\sigma_0 \xi g\eqno(C41)$$

To calculate the stress field inside the solid we need that as $\omega \rightarrow 0$ to leading order in $\omega$
$$p_T = i q \left (1- {\omega^2 \over c_T^2 q^2}\right )^{1/2} \approx iq - iq {1\over 2} {\omega^2 \over c_T^2 q^2}$$
so that
$$e^{i p_T z} \approx e^{-qz} \left (1+qz {1\over 2} {\omega^2 \over c_T^2 q^2}\right )\eqno(C42)$$
and similarly
$$e^{i p_L z} \approx e^{-qz} \left (1+qz {1\over 2} {\omega^2 \over c_L^2 q^2}\right )\eqno(C43)$$
For $z>0$ (C26) takes the form
$$-i\sigma_{xx} ({\bf q},z)= 2\mu q_x^2 A({\bf q})e^{i p_L z} -2\mu q_xq_y B({\bf q})e^{i p_T z}$$
$$- 2\mu p_Tq_x^2 C({\bf q})e^{i p_T z}+\lambda \left ({\omega \over c_L}\right )^2 A({\bf q})e^{i p_L z}\eqno(C44)$$
Substituting (C18)-(C20) and (C42) and (C43) in this equation gives as $\omega \rightarrow 0$:
$$\sigma_{xx}=  {q_x^2 \over q^2}  \left ({1\over c_T^2}- 
{1\over c_L^2} \right )^{-1} \left ({\lambda \over \mu c_L^2}+{1\over c_T^2} 
{q_x^2\over q^2} \right ) q h({\bf q})\sigma_0 e^{-qz}$$
$$ +2{q_x^2 q_y^2 \over q^3}  q h({\bf q})\sigma_0 e^{-qz} -{q_x^4\over q^4} qz q h({\bf q})\sigma_0 e^{-qz} $$
$$= \left [ 2 {q_x^2\over q^2} \left (1+\nu {q_y^2\over q^2} \right )
-{q_x^4\over q^4} qz \right ] qh({\bf q})\sigma_0 e^{-qz}\eqno(C45)$$
Thus
$$\sigma_{xx}({\bf x},z) = \sigma_0 \int d^2q \ qh({\bf q}) e^{i{\bf q}\cdot {\bf x}-qz}$$
$$\times \left [ 2 {q_x^2\over q^2} \left (1+\nu {q_y^2\over q^2} \right )
-{q_x^4\over q^4} qz  \right ] \eqno(C46)$$
In a similar way one can deduce the other components of the stress sensor $\sigma_{ij}$.
It is also interesting to calculate the ensemble average
$\langle \sigma_{xx}^2({\bf x},z) \rangle$.
From (4) and (9) it follows that
$$\langle h({\bf q}) h({\bf q}') \rangle = C({\bf q}) \delta ({\bf q}+{\bf q}')$$
Using this equation we get
$$\langle \sigma_{xx}^2({\bf x},z) \rangle =
\sigma_0^2 \int d^2q \ q^2 C({\bf q})  e^{-2qz}$$
$$\times \left [ 2 {q_x^2 \over q^2} \left (1+\nu {q_y^2\over q^2} \right )
-{q_x^4\over q^4} qz  \right ]^2 \eqno(C47)$$
For a system with isotropic roughness $C({\bf q})$ depends only on $q=|{\bf q}|$ and in that case the angular integration
in (C47) is easy performed giving
$$\langle \sigma_{xx}^2({\bf x},z) \rangle =
2 \pi \sigma_0^2 \int dq \ q^3 C(q)e^{-2qz}$$
$$\times \left [ {3\over 2} (1+\nu)^2-{5\over 4} (1+\nu) (2\nu+qz)+{35\over 128} (2\nu+qz)^2\right ] \eqno(C48)$$

As an illustration, if surface roughness occur only in the $x$-direction then $h({\bf q}) = h(q_x)\delta (q_y)$ so that
$q^2=q_x^2$ and
$$\sigma_{xx}=  \sigma_0 \int dq_x \ ( 2-q_xz) h(q_x) e^{iq_x x -qz}\eqno(C49)$$
which is the same result as obtained in Ref. \cite{d1,d2,d3}. 

It is easy to extend the analysis to the case where a uniform stress $\sigma^0_{yy}$ occur in addition to the stress $\sigma^0_{xx}$ denoted
by $\sigma_0$ above. Here we consider the particular simple case where $\sigma^0_{xx}=\sigma^0_{yy}=\sigma_0$.

Consider a rectangular block elongated in both the $x$ and the $y$-directions
with the same stress so that $\sigma^0_{xx}=\sigma^0_{yy}=\sigma_{0}$. 
In this case
$${\pmb \sigma} ({\bf x}) = (h'_x ,h'_y ,0) \sigma_0$$ 
and 
$${\pmb \sigma} ({\bf q}) = i {\bf q} h({\bf q})\sigma_0 \eqno(C50)$$
Using (A18)-(A20) in Ref. \cite{Persson2} the scalar fields $A$, $B$ and $C$ are given by
$$A({\bf q}) ={1\over \mu S} 2p_T q^2 h({\bf q})\sigma_0\eqno(C51)$$
$$B({\bf q}) =0\eqno(C52)$$
$$C({\bf q}) = {-1\over \mu S} \left ({\omega^2\over c_T^2}-2q^2 \right ) h({\bf q}) \sigma_0\eqno(C53)$$
We are interested in the $\sigma_{xx}$ stress component which can be written as in (C26).
Substituting (C51)-(C53) in (C26) gives
$$-i\sigma_{xx}= {2p_T\over S} \left ({\lambda \over \mu} \left ({\omega \over c_L}\right )^2 q^2 + 
\left ({\omega \over c_T}\right )^2 q_x^2 \right ) h({\bf q}) \sigma_0$$
Using that (C23) we get 
$$-i\sigma_{xx}= -i q 
\left ({1\over c_T^2}- {1\over c_L^2} \right )^{-1} 
\left ({\lambda \over \mu c_L^2} + {1\over c_T^2} {q_x^2\over q^2}\right ) h({\bf q}) \sigma_0$$
Using (C16) this equation gives
$$\sigma_{xx} ({\bf q}) =2 q  h({\bf q}) \sigma_0 {\nu q_y^2 +q_x^2 \over q^2}\eqno(C54)$$
By symmetry
$$\sigma_{yy} ({\bf q}) =2 q  h({\bf q}) \sigma_0 {\nu q_x^2 +q_y^2 \over q^2}\eqno(C55)$$
Note that the average 
$${1\over 2} \left [ \sigma_{xx} ({\bf q})+\sigma_{yy} ({\bf q})\right ] = (1+\nu) q  h({\bf q}) \sigma_0$$ 
Note that if $h({\bf q}) =h(q_x) \delta (q_y)$ we get
$$\sigma_{xx} ({\bf q}) = 2 q_x h(q_x) \sigma_0 \delta (q_y)$$
$$\sigma_{yy} ({\bf q}) = 2 q_x h(q_x) \nu \sigma_0 \delta (q_y)=\nu \sigma_{xx} ({\bf q})$$

Finally, I note that in an earlier version of this paper which was published on Research Gate
an error was made in deriving the relation between the stress $\sigma_{ij} ({\bf q})$ and $h({\bf q})$.
The equations for the stress given in the original paper obey the correct boundary conditions
and the stress tensor obey the correct equation $\sigma_{ij,j}=0$ for force equilibrium,
but the solution does not satisfy the stress compatibility equations (Beltrami-Michell equations;
if the compatibility equations are violated there exist no displacement field which gives the strain 
or stress tensor obtained). 

\vskip 0.3cm
{\bf Appendix D: Stress probability distribution}

Here we calculate the probability distribution (13) for the stress $\sigma({\bf x})=\sigma_{xx}({\bf x})-\sigma_0$.
Since $\sigma ({\bf q}) = 2 \sigma_0 q h({\bf q})$ where $h({\bf q})$ is assumed to be a Gaussian random variable
so will be $\sigma ({\bf q})$ and hence $\sigma ({\bf x})$. Using this we get
$$P(\sigma) = \langle \delta (\sigma - \sigma({\bf x}) ) \rangle 
= {1\over 2 \pi} \int_{-\infty}^\infty d\alpha \ \left \langle e^{i\alpha (\sigma - \sigma({\bf x}) )} \right \rangle $$
$$= {1\over 2 \pi} \int_{-\infty}^\infty  d\alpha \ e^{i\alpha \sigma} \left \langle e^{-i\alpha \sigma({\bf x})}\right \rangle $$

$$= {1\over 2 \pi} \int_{-\infty}^\infty  d\alpha \ e^{i\alpha \sigma -\alpha^2 \sigma^2_{\rm rms}/2}\eqno(D1)$$
where
$$\sigma^2_{\rm rms} = \langle \sigma^2 ({\bf x}) \rangle$$
In deriving (D1) we have used that for a Gaussian random variable the cumulant 
expansion is truncated at leading order. Performing the $\alpha$-integration in (D1) gives
$$P(\sigma) = {1\over (2\pi)^{1/2} \sigma_{\rm rms}} e^{-(\sigma/\sigma_{\rm rms})^2/2} \eqno(D2)$$

\vskip 0.3cm
{\bf Appendix E: Spatial stress distribution}

The analysis in Appendix C [see (C45)] shows that the stress field from a surface roughness components 
with wavenumber $q$ decay into the solid as $(a+bz){\rm exp} (-qz)$ where $a$ and $b$ depends on the
elastic properties of the solid. The exponential decay follows if the displacement field would obey 
a Laplace-type of equation. Thus the solution to
$$\nabla^2 u = 0$$
which vary as ${\rm cos}({\bf q}\cdot {\bf x})$ parallel to the surface, is of the form
$\sim {\rm cos}({\bf q}\cdot {\bf x}) {\rm exp} (-qz)$. The additional factor $(a+bz)$ in the actual
stress distribution is due to the fact that in the elastostatic limit the displacement field obey a 
biharmonic type of equation rather than the Laplace equation.

The exponentially decay of the stress field into the solid from each wavelength components of the roughness
is consistent with the Saint-Venant's Principle which state that the way the
loads are applied only matters for the stress field 
close to the point (or here the surface) of application\cite{Ven1,Ven2,Ven3}. 
Thus, a short distance
(here the wavelength of a roughness component) from the applied load the stress becomes uniform; in
our case it must vanish as the total normal force from a roughness component vanish [it oscillates as
${\rm cos} ({\bf q}\cdot {\bf x})$ parallel to the surface].

\vskip 0.3cm
{\bf Appendix F: Plasticity length} 

The plasticity length $d_{\rm Y}$ ranges from a few nanometers in some amorphous solids, 
to several micrometers or more in metals.
The energy per unit area to break bonds between atoms in solids is of order $E a$ where $E$ is the 
Young's modulus and $a\approx 0.2 \ {\rm nm}$ 
a bond distance. This follows from the fact that a strain of order 1 result in the the elongation of the 
bonds between the atoms by a factor of $\sim 2$ which is of order the distance needed to break an atomic bond. 
More accurately, if we write $\gamma = \alpha E a$ then experimental data and theory gives $\alpha \approx 0.1$. 

The yield stress varies strongly on the solid\cite{yield}. Plastic deformations are
stress aided, thermally activated processes and hence 
depend on the temperature (and the strain rate), and here we assume room temperature\cite{yield1}.  
For amorphous solids (e.g. silica glass) plastic 
deformation involves local rearrangements of the atoms in nano-sized volume elements\cite{amor}. The stress needed for 
the local atomic rearrangements is smaller than the stress to break the bonds, because plastic
yield events involve simultaneous bond-breaking and bond-formation and
require less energy (and less force or stress) than needed to separate the atoms completely. 
If we write $\sigma_{\rm Y} = \beta E$, then $\beta < \alpha$. 
Thus soda-lime (silica) glass, fused silica and amorphous silicon have $\beta \approx 0.03-0.05$.  

The plastic yielding in crystalline materials usually involves dislocations, and is 
fundamentally different from in the corresponding amorphous state.
The plastic yield stress is usually smaller in the crystalline state, but for some non-metallic systems the difference is small,
e.g., fused silica (amorphous ${\rm Si} {\rm O}_2$)
has $\beta \approx 0.04$, while quartz (crystalline ${\rm Si} {\rm O}_2$) has $\beta \approx 0.03$. Similarly sapphire 
(crystalline ${\rm Al}_2{\rm O}_3$) has $\beta \approx 0.02$. The similarity of the $\beta$ parameter for the amorphous and crystalline
state of some (non-metallic) solids indicate that the stress needed to move dislocations (the so called Peierls stress) in these materials
is similar to the stress needed to induce the local atomic rearrangements involved in plastic deformation of the amorphous state.  

For many metals the bond energy depends only weakly on the 
the detailed spatial (angular) arrangements of the atoms, assuming bond length are unchanged.
This is supported by the success of the jellium model (where the ions are smeared 
out into a uniform positive charged background) in describing
many properties of ``simple'' metals (e.g., the alkali metals and aluminum)\cite{jellium}.
In these cases even a small external stress may result in a rearrangement of the atoms.
Thus for crystalline metals slip of atomic planes over each 
other occurs at relatively low applied stresses, and plastic flow involves
movement of dislocations. Hence for metals $\beta$ is very small, e.g. $\beta \approx 5 \times 10^{-4}$
for pure aluminum and iron, and even for the hard material tungsten $\beta$ is relative small,
$\beta \approx 4 \times 10^{-3}$. For alloys the yield stress is higher 
than for the pure metals because the alloy atoms result in energetic barriers for the motion of 
dislocations. Thus for steel and aluminum 
alloys typically $\beta \approx (1-4)\times 10^{-3}$. Using that
$$d_{\rm Y} = {E \gamma \over \sigma_{\rm Y}^2} = {\alpha \over \beta^2} a $$
we get $d_{\rm Y} \approx 10 \ {\rm nm}$ for amorphous silicon or silica\cite{plast}, 
and $\approx 10 \ {\rm \mu m} $ or more for metals.

That metals are plastically much softer than materials like silica may be related to the electronic band structure.
Metals have no band gap and the response of the electrons to small displacement of the ions or atoms can be described in
perturbation theory as involving (virtual) low-energy excitation's (electron-hole pairs close to the Fermi surface), 
while in solids with wide band gaps, such as quartz (crystalline silica), 
the lowest energy excitation's have very large energies. In the latter case
we expect a larger energy barrier for atom rearrangements. 

In metals the atoms have many neighbors forming close-packed structures such as 
face-centered-cubic or body-centered-cubic structures, as expected from the closest packings of spheres.
Using a simple real space tight binding electronic structure model\cite{Hein} one can show that 
for metals the binding energy is proportional to the square-root of the number of nearest neighbors.  
This imply that creating local defects involving a slight change in the number of nearest neighbors is energetically
cheap and also that the shear modulus $G$ is smaller than expected if the binding 
energy would be proportional to the number of nearest neighbors\cite{Mus}. 
(Note: In simple models the elastic energy of dislocations is proportional to $G$.)
Thus for metals one expect the energies for atom rearrangements to be 
small as long as there are only small local changes in the
atom density and the number of neighbors. This simple model also provide insight in cases where the number of neighbors change,
including surface energies, stacking fault energies, energies of surface steps and more.


\begin{thebibliography}{99}
\bibitem{SK}
The stress concentration factor is usually denoted by $K$ but we use the notation $S$
in order not to confuse it with the stress intensity factor $K$ used in the theory of cracks.
\end{thebibliography}
\end{document}